\documentstyle[epsfig]{aipproc}

\begin{document}

\title{A Survey of the Host Galaxies of Gamma-Ray Bursts\footnote{Based on
      observations with the NASA/ESA {\sl Hubble Space Telescope}, obtained
      at the Space Telescope Science Institute, which is operated by the
      Association of Universities for Research in Astronomy, Inc. under
      NASA contract No. NAS5--26555.}}

\author{Stephen Holland$^*$}
\address{$^*$Department of Physics \\
             University of Notre Dame \\
             Notre Dame, IN 4655--5670 \\
             U. S. A.}

\maketitle


\begin{abstract}
     We used 45 {\sl HST\/}/STIS orbits in Cycle 9 to obtain deep
images of the host galaxies of eleven gamma-ray bursts.  Our goals are
to study the morphologies of the host galaxies, to obtain precise
locations of gamma-ray bursts within their host galaxies, and to
determine star-formation rates in the hosts.  We present preliminary
results for GRB~980425/SN1998bw, GRB~980613, and GRB~9809703.

\end{abstract}


\section*{Introduction}

     We have obtained deep images of the host galaxies of eleven
gamma-ray bursts (GRBs) using the Space Telescope Imaging Spectrograph
(STIS) aboard the {\sl Hubble Space Telescope\/} ({\sl HST\/}).  Data
was taken using the 50CCD (clear) and F28X50LP (long pass) apertures,
which peak at approximately the $V$ and $R$ photometric bands.  Each
image was taken more than one year after the burst to study the
overall morphology, and the small-scale structure of each host,
without contamination from the GRB's optical afterglow (OA).  Our
goals are to classify the morphologies of the host galaxies, identify
substructure in each host, and to probe the star-formation rate at
high redshifts.  Combining ground-based observations of the OAs with
our {\sl HST\/} data allows us to determine precise positions for each
GRB relative to substructure (such as bulges, spiral arms, H{\sc ii}
regions, star-forming regions, etc.) in the host.  The images will be
used to compare the distribution of host morphologies to galaxies in
the Hubble Deep Fields, and to search for correlations between
specific types of substructure and GRBs.  Morphological information,
and the spectral energy distribution, will allow us estimate overall
star-formation rates and the amount of dust present.  We have waived
all proprietary rights to this data. The reduced data are available at
{\tt http://www.ifa.au.dk/{\~{}}hst/grb\_hosts/index.html}.  This
paper presents some preliminary results from this project.


\section*{GRB~980425}

     There is growing evidence that GRBs are related to the deaths of
massive stars.  GRB~980425 and the Type Ib/c supernova SN1998bw
coincided in time and position on the sky \cite{sholland:GVvP1998},
GRB~990123 coincided with a star-forming region
\cite{sholland:HH1999}, and there is evidence for a GRB/SN connection
for eleven other GRBs (see \cite{sholland:HFH2001} for a list).

     The host galaxy of GRB~980425 is ESO~184$-$G82, a sub-luminous,
barred spiral (SBc) galaxy that is in a stage of strong star
formation.  The galaxy lies at a distance of 36.64 Mpc, making
GRB~980425 the nearest known GRB.  The host has a size, structure,
luminosity, and star formation rate similar to those of the Large
Magellanic Cloud.  ESO~184$-$G82 appears to be a member of a compact
group with the nearest neighbour located at a projected distance of
only 11.9 kpc and shows indications of being morphologically
disturbed.  Therefore, this galaxy may be undergoing
interaction-induced star formation.

\begin{figure}[!hb]
\begin{center}
\epsfig{file=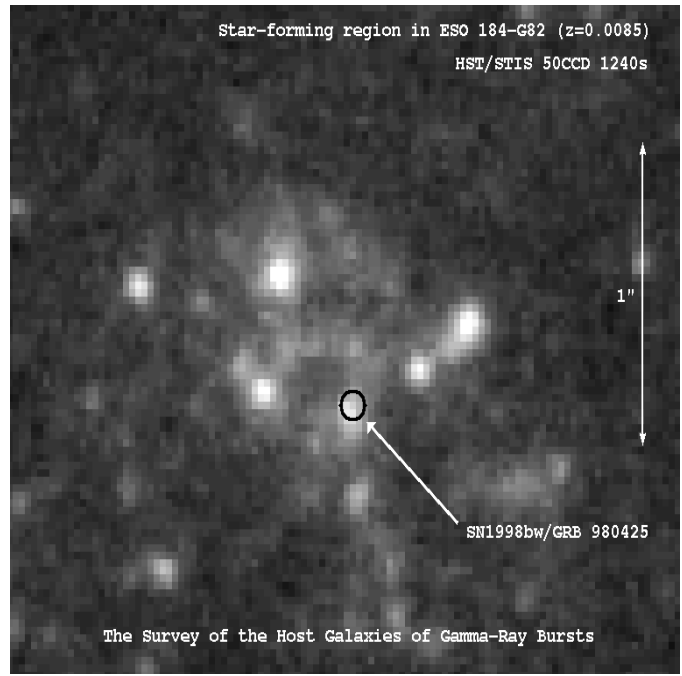,height=3.5in,width=3.5in}
\end{center}
\vspace{10pt}
\caption{This Figure shows our {\sl HST\/}/STIS 50CCD image of the
star-forming region where GRB~980425/SN1998bw occurred.  The plate
scale for all the {\sl HST\/}/STIS images presented in this paper is
$0\farcs0254$/pixel.  The circle shows the location of the
supernova/GRB.  Three of the stars in the star-forming region have
colours that are consistent with red giants while three stars have
blue colours that are consistent with their being massive
main-sequence stars \protect\cite{sholland:FHA2000}.}
\label{F:sholland:grb980425}
\end{figure}

     SN1998bw/GRB~980425 occurred in an H{\sc ii} region approximately
$300$ pc in diameter.  There are several bright, young stars within a
projected distance of approximately $100$ pc of the supernova/GRB
(Fig.~\ref{F:sholland:grb980425}).

\section*{GRB~980613}

\begin{figure}[hb]
\begin{center}
\centerline{\epsfig{file=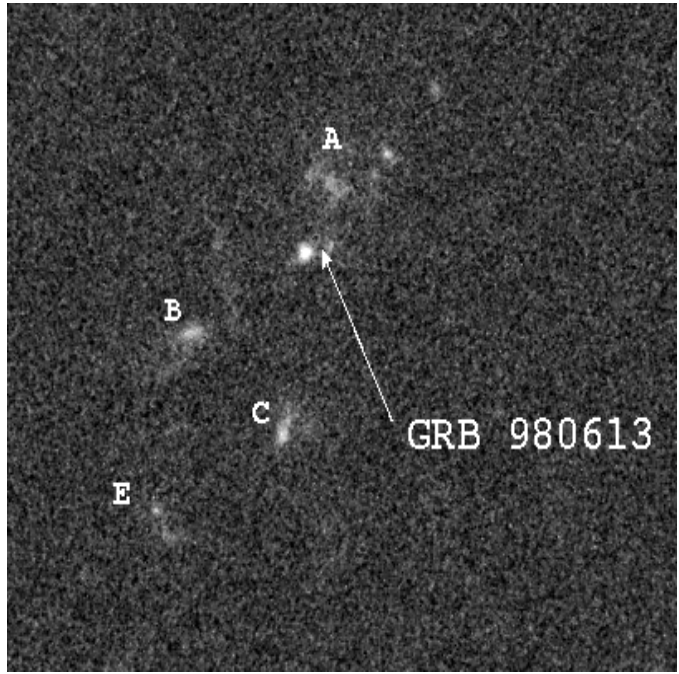,height=3.5in,width=3.5in}}
\end{center}
\vspace{10pt}
\caption{This is our STIS 50CCD (clear) image of the environment of
GRB~980613.  The resolution (i.e., the apparent diameter of a point
source) is $0\farcs084$.  The field of view is approximately $7\farcs5
\times 7\farcs5$ ($= 60 \times 60$ kpc assuming
$(H_0,\Omega_m,\Omega_{\Lambda}) = (70,0.3,0.7)$).}
\label{F:sholland:grb980613}
\end{figure}

     GRB~980613 occurred near the edge of a compact blue star-forming
region which may be part of a larger structure
(Fig.~\ref{F:sholland:grb980613}).  The morphology of the host is
chaotic, and we find no evidence for spiral structure, or faint
substructure connecting the various components.  We find a
star-formation rate of $\approx 3 {\cal M}_{\sun} {\rm yr}^{-1}$
assuming no extinction in the host galaxy.  The rest-frame $B$-band
luminosity is $L_B \approx (0.2 \pm 0.1) L^*_B$ where $L^*_B$ is the
luminosity of a typical galaxy at a redshift of $z \approx 1$.  The
specific star-formation rate per unit blue luminosity is $\approx 20$
${\cal M}_{\sun} {\rm yr}^{-1} {L^*_B}^{-1}$, the highest value of any
known GRB host galaxy \cite{sholland:HAH2001}.

    Djorgovski~et~al.~\cite{sholland:DBK2001} suggested that the host
is a set of interacting galaxies.  If this is the case then the lack
of tidal features suggests that the system has only recently
interacted.  Alternatively, the colour and morphology of the host are
similar to what is seen in low surface brightness (LSB) galaxies.
However, the host galaxy shows strong nuclear activity.  This is
unlike LSB galaxies, which tend not to exhibit nuclear activity,
although some of the larger LSB disk galaxies do show nuclear activity
\cite{sholland:BIM1997} similar to that seen in the host of
GRB~980613.  The large specific star-formation rate is unusual for an
LSB.  However, most of the star formation occurs in the nucleus and is
not distributed through-out the galaxy.  Therefore, the host may be an
LSB galaxy where star formation is in the process of turning on.

\section*{GRB~980703}

\begin{figure}[!hb]
\begin{center}
\epsfig{file=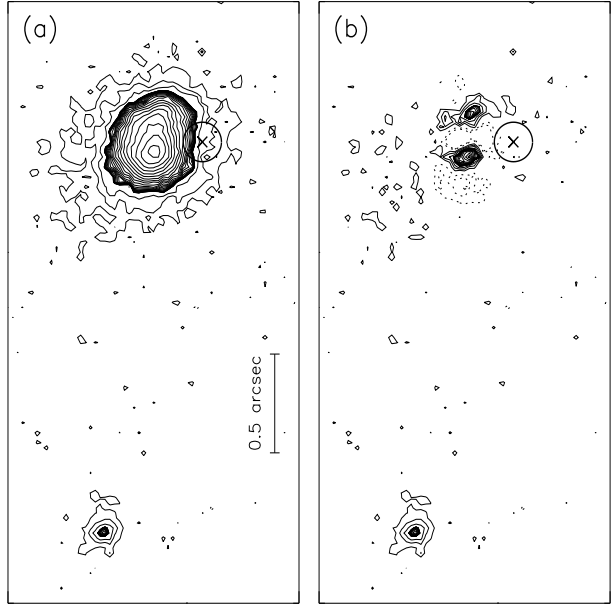,height=3.5in,width=3.5in}
\end{center}
\vspace{10pt}
\caption{{\bf a)}: This Figure shows the {\sl HST\/}/STIS 50CCD
GRB~980703.  The outer contours are linear while the inner contours
are logarithmic to show the host over a large dynamic range.  The
location of GRB~980703 is marked with an ``X'' and our 1$\sigma$ error
circle.  {\bf b)}: This Figure shows the host with the best-fitting
Sersic \protect\cite{sholland:S1968} model subtracted.}
\label{F:sholland:grb980703}
\end{figure}

     The host of GRB~980703 is a bright galaxy with an ``egg-shaped''
morphology that resembles a lop-sided barred spiral
(Fig.~\ref{F:sholland:grb980703}a).  We estimate a star-formation rate
of $8$--$13 {\cal M}_{\sun} {\rm yr}^{-1}$ using Eq.~2 of
\cite{sholland:MPD1998} and assuming that there is no extinction in
the host galaxy.  The rest-frame $B$-band luminosity is $(1.6 \pm 0.4)
L^*_B$.  Therefore, the star-formation rate per unit luminosity is
$\approx 6.5 {\cal M}_{\sun} {\rm yr}^{-1} {L^*_B}^{-1}$, which is
similar to the values found for several other GRB host galaxies.

     We fit two-dimensional Sersic \cite{sholland:S1968} models to the
{\sl HST\/}/STIS images, after convolving with the appropriate
point-spread function \cite{sholland:HFH2001}, and found a half-light
radius of $0\farcs13 \pm 0\farcs01$, $n = 1.05 \pm 0.02$ ($n = 4$
corresponds to a de~Vaucouleur $R^{1/4}$ profile), and an ellipticity
of $0.24 \pm 0.02$.  This is consistent with an exponential disc with
a scale radius of $0\farcs21 \pm 0\farcs01$.
Fig.~\ref{F:sholland:grb980703}b shows the host with the best-fitting
model subtracted.  Except for the central regions the galaxy is well
fit by this model.  The systematic residuals in the central few pixels
suggest that there is substructure in the galaxy.  The excess of light
on the west side of the galaxy may be a spiral arm.  The derived
half-light radius is much smaller than those seen in local late and
early type galaxies \cite{sholland:ICG1995}.  However, the size,
colour, and spectrum of the host are similar to those of compact
galaxies in the Hubble Deep Field North
\cite{sholland:PGG1997,sholland:GGK1997}.

\section*{Star Formation and Host Morphology}

     There is evidence that redshifts of GRBs can be determined from
the intrinsic properties of the bursts
\cite{sholland:S2001,sholland:RLF2001}.  These techniques will allow
the distribution of GRBs to be mapped to high redshifts solely from
the observed gamma-ray pulse.  However, in order to connect the GRB
rate with star formation in the early Universe it is important to
understand the connection between GRBs and star formation in
individual galaxies.  Preliminary results from our Cycle 9
observations suggest that long-duration ($T_{90} > 2$ s) GRBs are
located in star-forming regions
\cite{sholland:H2000,sholland:HFH2001,sholland:HAH2001}.  This is
consistent with the growing evidence that GRBs are produced in the
early stages of a supernova explosion.  Our data suggests that hosts
tend to be sub-luminous with high rates of star-formation per unit
luminosity (see Table~\ref{T:sholland:ssfr}).

\begin{table}[!hb]
\begin{center}
\caption{Specific star-formation rates for several GRB
host galaxies.}
\label{T:sholland:ssfr}
\begin{tabular}{lccd}
\multicolumn{1}{c}{GRB} &
\multicolumn{1}{c}{$z$} & 
\multicolumn{1}{c}{$R_{\mathrm{host}}$} & 
\multicolumn{1}{c}{${\cal M}_{\sun} {\rm yr}^{-1} {L^*_B}^{-1}$} \\
\tableline
970508 & 0.835 & 25.20 & 11.0 \\
980613 & 1.096 & 24.56 & 20.0 \\
980703 & 0.966 & 22.57 &  6.5 \\
990123 & 1.600 & 24.07 & 11.0 \\
990712 & 0.434 & 21.91 & 4.4 \\
\end{tabular}
\end{center}
\end{table}

\begin{figure}[!ht]
\begin{center}
\epsfig{file=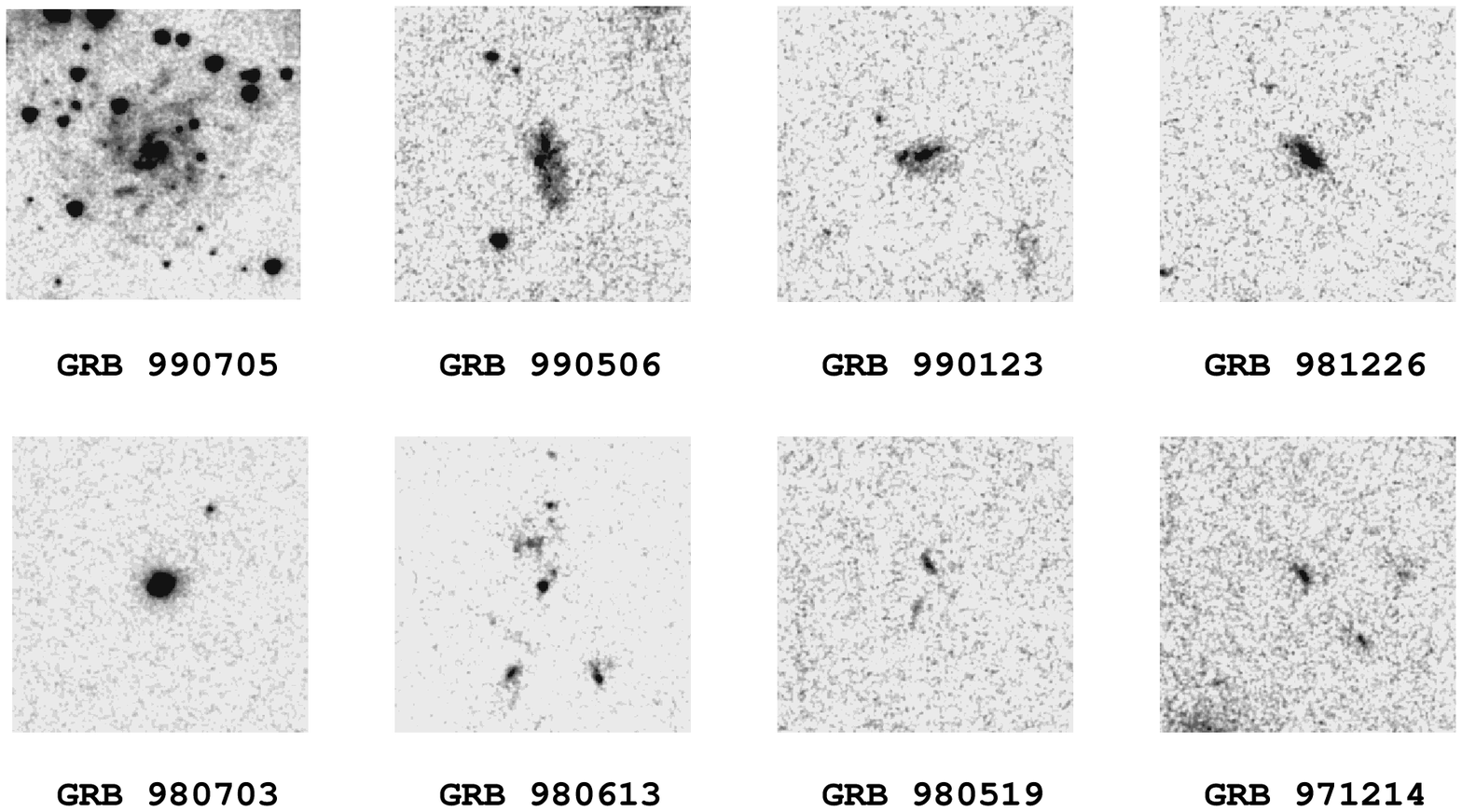,height=3.5in,width=6in}
\end{center}
\vspace{10pt}
\caption{This Figure shows {\sl HST\/}/STIS images of eight host GRB
galaxies.  Each image is in 50CCD except GRB~990123, which is in
F28X50LP.  The fields of view are $6\farcs5 \times 6\farcs5$.}
\label{F:sholland:gallery}
\end{figure}

     Fig.~\ref{F:sholland:gallery} shows images of eight GRB host
galaxies.  There is no single morphological type for GRB hosts,
although most appear to occur in late-type and irregular galaxies (but
see \cite{sholland:BSD2001}).  Our preliminary results suggest that
the morphologies of GRB host galaxies are no different from other
star-forming galaxies at the same redshift.

\end{document}